# Small Data and Process in Data Visualization: The Radical Translations Case Study

Arianna Ciula*  Miguel Vieira[†]  Ginestra Ferraro[±]  Tiffany Ong[§]  Sanja Perovic[¶]  Rosa Mucignat[≠]  Niccolò Valmori[%]  Brecht Deseure[$]

King's College London

Erica Joy Mannucci[&]

Università Milano-Bicocca

**ABSTRACT**

This paper uses the collaborative project *Radical Translations* [1] as case study to examine some of the theoretical perspectives informing the adoption and critique of data visualization in the digital humanities with applied examples in context. It showcases how data visualization is used within a King's Digital Lab project lifecycle to facilitate collaborative data exploration within the project interdisciplinary team – to support data curation and cleaning and/or to guide the design process – as well as data analysis by users external to the team. Theoretical issues around bridging the gap between approaches adopted for small and/or large-scale datasets are addressed from functional perspectives with reference to evolving data modelling and software development lifecycle approaches and workflows. While anchored to the specific context of the project under examination, some of the identified trade-offs have epistemological value beyond the specific case study iterations and its design solutions.

## 1 INTRODUCTION

Before dwelling on the details of the case study at hand, it is important to reflect briefly on the meaning of scale in the context of digital humanities projects of historical nature – i.e., focused on objects or phenomena of past cultures and societies – along the following dimensions:

1. the relation between the analogue and the digital archive;
2. the calibration towards the perspective of analysis;
3. the pragmatic and opportunistic stance of project workflows.

With respect to point 1, the concept of scale varies extensively whether applied for example to the digital archive or to the analogue one. Often, in digital humanities projects, only a small fraction of the analogue archives of interest is available in digital form; further, within the digital archive, only a share might be used as input dataset to shape visualizations or other types of data exploration and analysis (figure 1). In alignment with critical documentation approaches around data experiments in social sciences and machine learning methods [2], recent projects have started making these differentiations explicit, revealing not only gaps in the digital archive but limitations in the relations between the digital and the analogue archive [3]. Talking about scale without this level of contextualization can therefore be misleading.

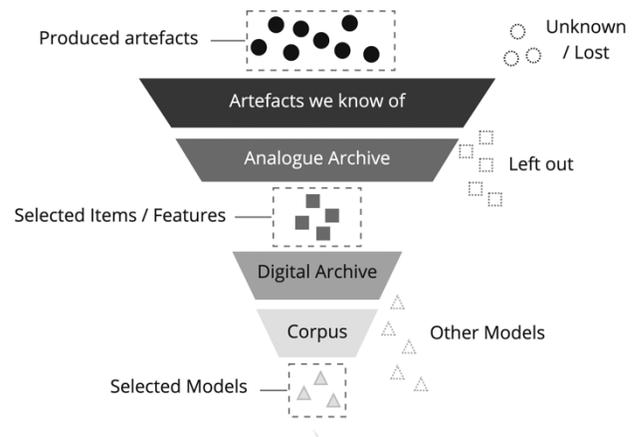

Figure 1: Process of reduction (selection and loss) of cultural artefacts from available items in the analogue archive to digital records and models.

The second dimension to consider relates to the calibration of scale with respect to the perspective of analysis deemed suitable for a specific research context or set of questions. Depending indeed on such perspective, a granular versus a bird's eye view on the data might be privileged. While interactive and storified, data visualizations can arguably support (or are conceived to support the interplay between) both modes of seeing [4], inevitably scale is calibrated to whatever perspective is foregrounded. A further element of calibration of scale deals with the implications of operationalizing [5] large and complex research concepts. Investing in data modelling to allow for scalability, replicability and transparency is therefore a conscious choice. While working on big data, this task is inevitable and count as a small percentage of time and resources against further development to yield meaningful and human readable results. In proportion this effort could seem to come at a higher cost for small(er) datasets.

* arianna.ciula@kcl.ac.uk
[†] jose.m.vieira@kcl.ac.uk
[±] ginestra.ferraro@kcl.ac.uk
[§] tiffany.ong@kcl.ac.uk
[¶] sanja.perovic@kcl.ac.uk
[≠] rosa.mucignat@kcl.ac.uk
[%] niccolo.valmori@kcl.ac.uk
[$] brecht.deseure@kcl.ac.uk
[&] erica.mannucci@unimib.it

Finally, scale is affected by very pragmatic factors in a project workflow. For example, in many digital humanities projects limited research funding is expected to cover both data collection (if not digitization *tout court*) and analysis with the consequence that data visualization might be squeezed into short and quick iterations with very limited availability of resources. This and other pragmatic factors inevitably reduce scale to opportunistic selections.

## 2 CASE STUDY

The project selected as case study is *Radical Translations: The Transfer of Revolutionary Culture between Britain, France and Italy (1789-1815)* [1], a collaboration between King's Digital Lab and the Departments of French and Comparative Literature at King's College London, and the University of Milan-Bicocca, led by Dr Sanja Perovic and funded by the Arts and Humanities Research Council (UK), 2019-22. The project's main objectives are to (i) provide a comparative study of the translation and circulation of democratic and free-thinking texts between Italy, France and Britain during the French Revolution and Napoleonic era; (ii) enhance public and academic awareness of the role of translation as an integral element of the revolutionary project; (iii) investigate how translation makes it possible for radical works to be 'living texts' that continually move forward into new communities, new places, new times. As technical partner in the project, King's Digital Lab worked with partners to define high level requirements at pre-project stage. This entailed the design and development of the overall technical solution for the project inclusive of a public website to make accessible relevant bibliographic and biographical data and within it, amongst other functionalities, five national timelines covering the three linguistic areas of the project (French, English and Italian) that showcase co-occurrences of macro-events relevant to both the history of radicalism as well as translation.

With respect to the relation between the analogue archive of relevance to the project and the digital archive, a few details and considerations need to be made. Firstly, the corpus for the project is small compared to other existing archives (both digital and analogue). This is partly due to the unusual focus not on the circulation of revolutionary-era translations per se (something that can be located using existing library catalogues) but on translations that seek to extend revolutionary ideas into new contexts. What counts as a radical translation and where and how it is found implies an interpretative framework and criteria of selection defined by the research team. In addition, an untold number of translations and fragments of translation appeared in newspapers, pamphlets and other ephemeral media, reaching a wider and more diverse readership than book circulation alone. A major challenge of this project is to recover this rich vein of revolutionary translations, often inserted without attribution and not registered in standard library catalogues. There are, however, interesting remarks to be made about the layers of scale with respect to the analogue archives which are reflected on the politics of digitization; for example, the Italian analogue landscape of bibliographic resources relevant for the project reflects the variety and fragmentation of political entities that characterized Italian history. This means that analogue sources are widely spread in different archives located in different cities and consequently separate (smaller) digital repositories. On the contrary, French and English sources, especially books and pamphlet literature, were easily traceable either via the respective national libraries or national archives digital catalogues (e.g., British Library and Gallica). The challenge was to find unattributed or acknowledged fragments of translation gleaned from publishers' prospectuses, newspapers, government archives and personal correspondence. Historically speaking, the fragmentation of the Italian analogue sources has clearly hindered the development of 'mass' digitization programmes at a regional or national scale. Yet foreign libraries (e.g., in Europe and the US) have managed to digitize Italian works as part of their effort of making their (bigger) collections available to a larger public. This phenomenon of circulation of the digitized versions of these works echoes the inherent circular vocation of their analogue counterparts that through numerous passages became part of a recognized global heritage.

With respect to the scope and research aims of the project that affected the scale of the dataset, a complex concept — namely transnational, shared European heritage of 'radicalism' as expressed in translation activities – had to be operationalized i.e., in this case distilled, reduced and abstracted out to be represented in a relational data model. Intellectual history often treats the circulation and reception of political texts and language over a long timescale. This project, through a focus on translation, offers a novel way of tracking the mobility of revolutionary language as it changes over a short timescale (1789-1815). The analysis of bibliographic metadata is therefore complemented in the project by close reading of paratexts (such as prefaces, addenda, titles, dedications) as opposed to full texts (which would have offered a very different scale). As typical of any data modelling iterations, the process of devising classification schemes compelled the project team to 'break down' complex (and at times confused!) concepts into constitutive blocks that could be used as descriptors and filters (e.g., paratext terms). Some of these descriptors are based on the material structure of the documents under study (e.g., whether a text has a preface or footnotes or whether it is an abridged, partial or new translation); others are function-terms which depend on the research team's interpretation of the communicative function of the paratext, i.e., whether it is mainly intended to build a community, explain a text, clarify culture-specific references etc. Therefore, early in the design process and requirements elicitation phase, it emerged that a granular view offering the opportunity for this analytical reading of metadata (as opposite to the full texts of the translations) had to be privileged.

One pragmatic element to highlight related to the bibliographic data available to the project at the start concerns data collection. A small share of the data was made available by the British Library as part of their French Revolution Tracts collection [6] under a CC0 licence to encourage research into the collection. However, given that project focus on 'radical' translations, this metadata could not be used wholesale; not only it had to be mapped and converted to the project choice of data model, but in fact also needed to be trawled through manually, like any library catalogue. It provided a slight advantage compared to other metadata that needed to be collected via *ad hoc* online searches and archival visits *in situ*. In the end, it was the knowledge of people ('radical authors or translators') that led to the identification of the texts that they may have translated. This 'manual' data collection process had to be accommodated within the first 2 years of the project during a global pandemic and therefore affected the scale of the dataset at hand, and it is still ongoing.

At the time of writing, the scale of the dataset in term of database records amounts to the following:

- 1288 agents (412 organizations, 876 persons, of which 254 are anonymous);
- 216 events;
- 1640 resources (of which 673 are translations and 224 paratexts).

## 2.1 Bridging Small and Large

With all the premises outlined above, the case study at hand would be classified as small and 'smart' [7] data. The data modelling perspective and technical architecture adopted for the project is cognizant not only of standards in the field that could make the dataset interoperable or at least amenable to further re-use and extension, but also to data exposure via APIs and Linked Open Data approaches that could enable multi-scale visualizations in the future (e.g. to showcase how many of the translators mentioned in the corpus under examination also appear on Wikidata or are invisible from other open data resources). The potential for smart bigger data is therefore not actualised but embedded in the model.

An overview of the project data model is available at [8]. While the data model could support data visualizations at different scale, its *raison d'etre* is principally to address the research questions outlined by the project team and by an interdisciplinary research team of intellectual historians and literary scholars. The role of selections and subjectivity are to be foregrounded, not only to give sense of the scale of the dataset and related visualizations as mentioned above, but also of its limitations. With this respect, the visualizations – some of which are reproduced below – generated for data exploration purposes at an early stage in the project lifecycle are useful to pinpoint the scale of the project (e.g., along the time axis) and highlight the subjective framing of the research space (e.g., top-down classification of types of bibliographic resources versus factual metadata).

A first data exploration exercise was developed using a Jupyter Notebook [9] and the Altair [10] library for data visualization. A second data exploration exercise was developed using both the Plotly [11] and Altair libraries.

Dynamic notebooks of this kind, developed to perform typically with large scale data are increasingly used for data exploration and analysis in the (digital) humanities; in the context of the small dataset, we outlined above, their function is not so much to reduce the space of features or identify modelling predictions but rather to serve the objectives of:

1. facilitating data checking and cleaning at early and mid-project stages;
2. defining (sometimes unexpected or existing) requirements with respect to interactive and more refined data visualizations; in this case:
    i. for the intellectual and literary historians to do analysis of the corpus or dataset under study;
    ii. and to expose it (mainly in the form of a timeline) on the project public website;
3. gain a better understanding of the data and potential new insights.

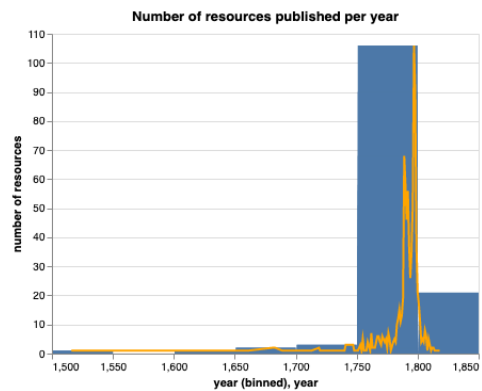

Figure 2: Number of resources published per year – it visualizes the chronological period range of interest to the project.

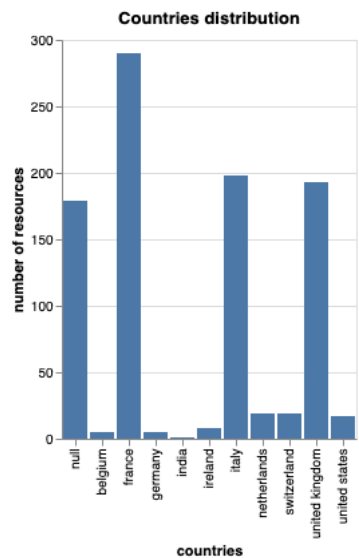

Figure 3: Country distribution – it visualizes the geographic selection defined by the project team from the outset. This also conditioned the project timeline discussed below. Note the null values relevant for data checking and curation.

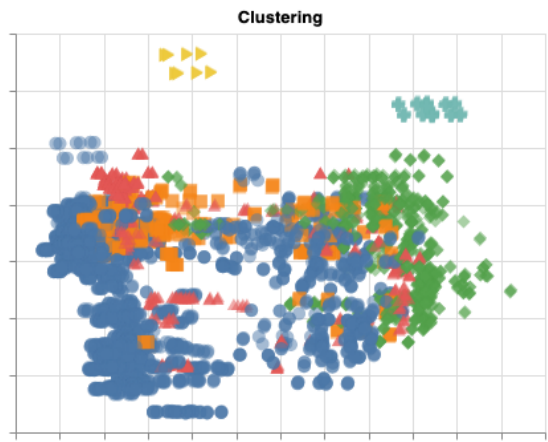

Figure 4: Clustering experiment – this experimental data exploration was not developed further at the time of writing.

## 2.2 Small Data and Processes

With a handful of exceptions, KDL projects deal with small data tailored to *ad hoc* research questions and contexts across its infrastructure (the lab currently maintains around seventy project servers and a total of 111 servers). The Software Development Lifecycle (SDLC) via which these projects are managed is however a rather homogenous process adapted (and in continuous improvement) from existing best practices in industry to the lab socio-technical context [12, 13].

Below we walk through a series of data visualizations iterations in relation to point 2.a above which evolved into the timeline [14] accessible on the project public website. These timelines (one for each country) were carefully constructed to reflect events that matter for translation, including censorship, regime change, military occupation and so forth. They also, importantly, can be compared to each other.

Some key points of the design process are summarized below and mapped to the figures and the role the different visualizations played in the SDLC evolutionary development phase:

1. when in front of first exploratory sketches with scatter plot format, it was agreed that not only political events but also translation events (i.e., bibliographic resources) would need to be plotted in the timeline – figures 5, 6;
2. prototypes using other visualization formats followed with variable levels of success in terms of readability factors (that could not be mitigated with changes in style) and unsuitability to the structure and distribution of the dataset – figures 7, 8, 9 ;
3. notebook to explore 2 formats more in detail in collaboration with partners (namely scatter plot and Gantt chart formats) – figures 10, 11;
4. convergence towards the grid format (a third option which emerged as part of further divergent explorations) and successive iterations to refine the design – figures 12, 13;
5. user testing outside the project team (scheduled for autumn 2021).

The list above is a *post-factum* selection; other libraries and tools were tested to explore other formats and inspire the team to diverge before converging on a selected option as per the double diamond approach [15] e.g., the heatmap plot was tested as well as network visualizations extending data points (nodes) beyond events and resources to persons and organizations.

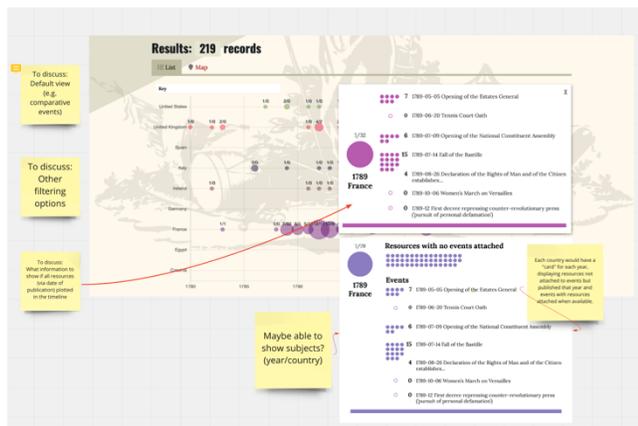

Figure 5: Example of use of a Miro [16] board to discuss remotely and collaboratively information to visualize on the timeline (scatter plot format).

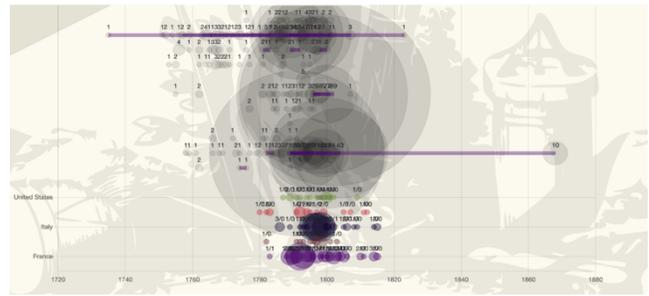

Figure 6: Iteration #1 plotted in the scatter plot format resulting in loss of readability.

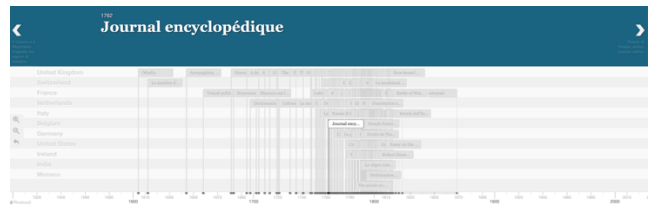

Figure 7: Test using TimelineJS [17] with problematic results in terms of information overlapping.

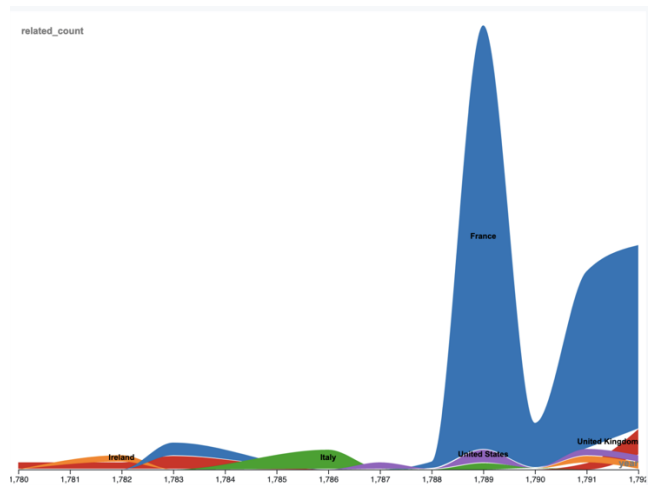

Figure 8: Test using Rawgraph [18] bump graph format on project dataset deemed unsuitable because of lack of enough detail.

The radial view is an interesting example of the second step mentioned above. It was tested to see if representing distribution in a circle would yield promising results, facilitating a bird's eye view and user interaction. However, the test with real data revealed it was an unsuitable visualization, because the radial view suggests cyclicity rather than foregrounding temporal linearity (important in this case). It also hindered the grouping and overlaps with all the relevant data points and features (date/event/resources/country/subjects). In turn, this realization helped choosing which features needed prioritizing ('resources' in this case).

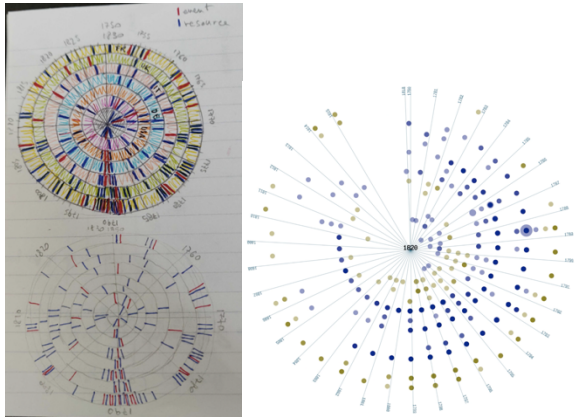

Figure 9: Radial view sketches and interactive tests designed adapting existing CSS code [19] but deemed unsuitable for the project dataset.

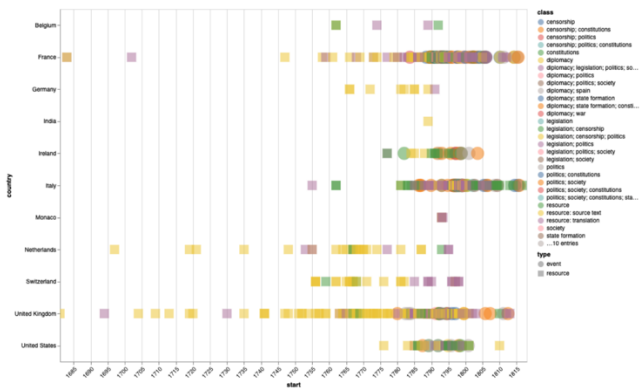

Figure 10: One of the versions of scatter plot format options discussed in one of the design iterations.

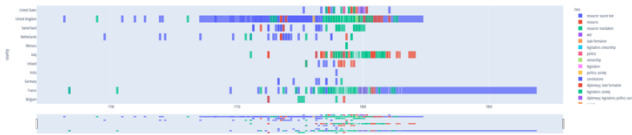

Figure 11: One of the versions of Gantt format option discussed in one of the design iterations.

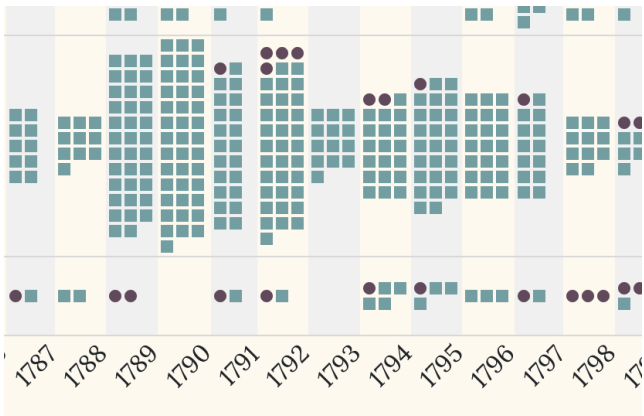

Figure 12: Grid view option discussed and selected in one of the design iterations. This view was inspired by KDL Solution Development Team explorations of Vikus viewer [20, 21, 22].

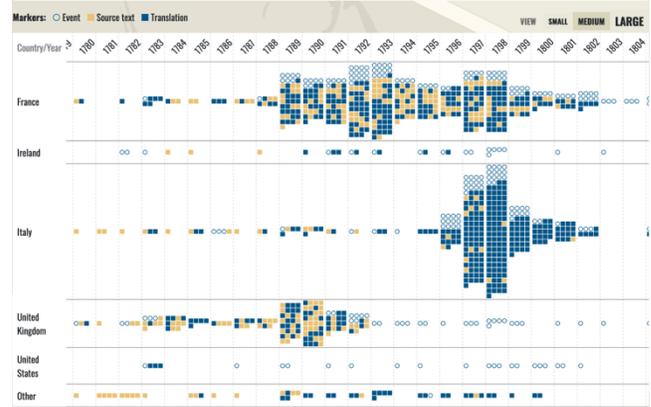

Figure 13: Final iteration prior to live deployment.

## 3 TRADEOFFS AND CONCLUSIONS

By combining the case study with the theoretical perspectives outlined above, we can complement the literature on data scholarship in the humanities [e.g., 23, 7] and identify at least three levels of trade-offs that data visualizations bring to the fore in digital humanities projects of small scale and processes.

First, in the humanities there seem to be an epistemological fallacy whereby visualizations (and digital methods more in general) are paired up with 'objective' quantification. On the contrary and by necessity, as the paper exemplifies, data visualization in the humanities is subjectively marked from data selection and collection onwards throughout the design process [24]; it is fallacious to assume that by creating data visualization, the inevitable element of subjectivity would be or need to be eclipsed. In collaborative projects engaged with knowledge infrastructures such as libraries, archives and laboratories, subjects are multiple and varied. We attempted to make emerge the interrelations of some of these subjectivities, from the KDL's team's input and choices to the historians' objective sand selections in the project; from the SDLC process being adopted to adapted libraries and platforms; from politics of digitisations to gaps in the archives.

Second, a sort of representational fallacy tends to occur. Adopting DH methods in visualizing data does not mean to embrace or surrender to quantification *tout court* but on the contrary, it means negotiating between visual languages – that are conventionally associated to predominantly quantitative arguments [25] – and the expressiveness of nuanced and subjectively selected and modelled data. While visual expressiveness gains from simplicity and reduction, this process of visual abstraction inevitably generates a trade off with the complexity of the dataset under examination, for example with respect to the underlying data model.

Finally, the effect of user engagement with project data at a glance is traded off with the necessity of a slow approach to data [26] that calls for detailed examinations of units of analysis, their meanings and relationships. Ultimately, the optimal balance of this tradeoff is context-dependent in that it is guided by the user interaction we are offering or aiming to offer. In other words, the level of detailed information and available actions the user gets from interacting with the data is assessed holistically with other design elements that complement the single visualization. In the case of the timeline showcased above, the combination of filters as a search tool and the actions on the "results", is key to fine tune how much information (direct and indirect) a visualization ought to hold and offer. User testing will be critical in assessing the overall design with this respect.